# Ground and Excited Electronic Structures of Electride and Alkalide Units: The Cases of Metal-Tren, -Azacryptand, and -TriPip222 Complexes

Isuru R. Ariyarathna

*Department of Chemistry and Biochemistry, Auburn University, Auburn, AL 36849-5312, USA*

Current address: *Physics and Chemistry of Materials (T-1), Los Alamos National Laboratory, Los Alamos, NM 87545, USA*

Email: ira0002@auburn.edu, isuru@lanl.gov

**Abstract**

A systematic electronic structure analysis was conducted for $M(L)_n$ molecular electrides and their corresponding alkalide units $M(L)_n@M'$ (M/M' = Na, K; L = Tren, Azacryptand, TriPip222; $n$ = 1, 2). All complexes belong to the "superalkali" category due to their low ionization potentials. The saturated molecular electrides display $M^+(L)_n^-$ form with a greatly diffuse quasispherical electron cloud. They were identified as "superatoms" considering the contours of populating atomic-type molecular orbitals. The observed superatomic Aufbau order of $M(Tren)_2$ is 1S, 1P, 1D, 1F, 2S, 2P, and 1G and it is consistent with those of M(Azacryptand) and M(TriPip222) up to the analyzed 1F level. Their excitation energies decrease gradually moving from $M(Tren)_2$ to M(Azacryptand) and to M(TriPip222). The studied alkalide complexes carry $[M(L)_n]^+@M'^-$ ionic structure and their dissociation energies vary in the sequence of $K(L)_n@Na > Na(L)_n@Na > K(L)_n@K > Na(L)_n@K$. Similar to molecular electrides, the anions of alkalide units occupy electrons in diffuse Rydberg-like orbitals. In this work, excited states of $[M(L)_n@M']^{0/+/-}$ and their trends are also analyzed.

## I. Introduction

Electrides are truly unique in the sense that electron localization is found in lattice voids rather than in the vicinity of cationic metal centers. Hence deservingly, electrides and its akin alkalides have attracted a notable attention over the years (see Refs. 1–5 and references therein). An electride is simply an outcome of metal atom coordination with particular ligands that prompts spontaneous metal-to-interstitial electron transfers.[1,2] On the contrary, in alkalides the anionic interstitial sites are populated by alkali metal atoms in the form of anions; i.e., Na$^-$, K$^-$, Rb$^-$, or Cs$^-$.[5,6] Electrons of alkalides are weakly bound to the crystalline framework similar to electrides due to the lower electron affinity of interstitial alkali metal atoms.[7] Such unique electron arrangements endow electrides and alkalides with properties such as low work function,[8] antiferromagnetism,[9] strong reducing power,[4,10–12] and high hyperpolarizability.[13,14] Hence, they are potential candidates in catalysis, optics, electronics, and magnetic materials related applications.[4,5,13,15,16]

Despite a variety of electrides and alkalides being discovered in ambient conditions, their experimental analysis bear challenges due to their low stability. Consequently, computational tools have been a preferred method to shed light on their properties.[3,4,17–21] Notably, such techniques are used to confirm the existence of interstitial electrons in electrides[17] and ion pairing in alkalide solutions.[10] Furthermore, theoretical studies of precursors of electrides and alkalides provide insight on their electronic structures enabling progress in designing, property tuning, and potential synthesizing of desirable electron rich materials.[10,22–25]

The building blocks of saturated organic electrides themselves are a fascinating class of complexes with unique electronic structures. For example, units such as Na(15-Crown-5), K(18-Crown-6), Na[2.2.2]cryptand, and K[2.2.2]cryptand possess a quasispherical highly diffuse electron cloud in the periphery (like an s-orbital) with an ionic structure.[22,23] Even more interestingly, these complexes populate atomic p-, d-, f-type orbitals (or superatomic (SA) P-, D-, F-orbitals) in excited states, and hence are identified as "Superatoms". Our past investigations disclosed their Aufbau shell model, which is chiefly 1S, 1P, 1D, 1F, 2S, and 2P (in some instances 2S populates prior to the 1F).[22,23] Similar orders are observed in a series of other Rydberg-like systems as well, namely, endohedral M@C$_{20}$H$_{20}$ (M = Li, Na, K, Rb),[26,27] $M(NH_3)_4$ (M = Li, Na),[28] and $M(H_2O)_6$ (M = $Mg^+$, $Ca^+$).[29,30] Complexes with such expanded electron clouds tend to carry extremely low ionization potentials and can also be recognized as "Superalkalis".[31,32]

As an extension to our previous efforts on organic electrides, here we have focused on ground and electronically excited states of several metal-coordinated Tris(2-aminoethyl)amine (Tren), Azacryptand (Azacrypt),[33] and TriPip222 cryptand (Tripip)[34] molecular electrides and their corresponding alkalide units intending to explore their lower ionization potentials and SA nature. Such electrides (built by N-based ligands) are more stable compared to electrides made with O-based ligands; e.g., Na(Azacrypt) vs. Na[2.2.2]cryptand.[35,36] Indeed, the focused Na(TriPip) is the first ever crystalline organic electride synthesized that is stable at room temperature.[35] Note that, due to the comparable ligand architecture and the identical coordination number, the studied $M(L)_n$ electride and $M(L)_n@M'$ alkalide systems (M/M' = Na, K; L = Tren, Azacrypt, Tripip; n = 1, 2) are expected to provide particularly useful information for fundamental property comparisons on these complexes and similar SA and superalkali entities. This article highlights their geometric features, chelation strengths, and electron ionizations and transitions. Furthermore, their SA nature originating from the shapes of the molecular orbitals and the smaller ionization potentials are discussed in detail. In this regard, their ground state geometries, dissociation energies, ionization energies (IEs), and excitation energies are obtained using DFT (density functional theory), MP2 (Møller–Plesset second-order perturbation theory), and EPT (electron propagator theory) and are discussed in the following sections.

## II. Computational details

The geometries of all reported species were optimized at the DFT/CAM-B3LYP[37] level using correlation-consistent double-ζ quality aug-cc-pVDZ basis set.[38,39] Cartesian coordinates and the harmonic vibrational frequencies of optimized complexes are reported in the Supporting Information (SI) (Tables S1–S13). At the DFT/CAM-B3LYP level, dissociation energies ($D_e$) of all the selected molecular electrides and their cations were calculated with respect to $M(L)_n^{0,+} \rightarrow M^{0,+} + nL$, when all fragments are at their ground electronic states (M = alkali metal, L = chelating ligand, n = 1, 2). The dissociation energies of the alkalide units ($D_e^*$) and their cations were calculated with respect to $[M(L)_n@M']^{0,+} \rightarrow [M(L)_n]^{0,+} + M'$ (M' = external alkali metal atom) fragmentation. At the same level of theory, adiabatic ionization energies (AIEs) of all species were collected. Utilizing the same basis set, single-point MP2[40] calculations were also performed for the CAM-B3LYP geometries of neutral and cationic Na(Tren)$_2$@Na, Na(Tren)$_2$@K, and K(Tren)$_2$@Na, to obtain more accurate $D_e^*$s and AIEs and to compare with CAM-B3LYP values. The spin contamination associated with the unrestricted Hartree–Fock wave functions of open-shell complexes is minor (less than 0.0003).

The vertical electron attachment energies (VEAEs) of M(Tren)$_{1,2}^+$, M(Azacrypt)$^+$, and M(Tripip)$^+$ were calculated using the EPT; i.e., Koopmans's theorem (KT), diagonal second-order (D2), partial third-order quasiparticle (P3), and renormalized partial third-order quasiparticle (P3+) methods,[41–45] using the previously benchmarked double-ζ quality cc-pVDZ (M, C, N) d-aug-cc-pVDZ (H) basis set. These EPT calculations were performed using the geometries of the neutral complexes. The excitation energies of the neutral complexes were inferred from the differences of the VEAEs of the cations.[43–45] The D2 VEAEs of $[M(L)_n@M']^{2+}$ and M(L)$_n$@M' were computed using the $[M(L)_n@M']^+$ and $[M(L)_n@M']^-$ geometries with the aug-cc-pVDZ basis set for all atoms, to analyze their low-lying electronic states. Furthermore, at the same level of theory vertical singlet-triplet energy gaps ($\Delta E_{S-T}$) of M(L)$_n$@M' were obtained at the geometry of singlet spin structures. For all EPT calculations CAM-B3LYP geometries were used. In all cases, pole strengths corresponding to the VEAEs are larger than 0.85 and, hence excitation energies are reliable. Note that the norms of Dyson orbitals corresponding to VEAEs are equal to the pole strengths (see Refs. 43–45 for more information). Calculated VEAEs and the corresponding pole strengths of all species are reported in the SI (Tables S14, S16, S18, S20, S22, S24, S26, S27, S28).

All calculations were performed using the Gaussian 16 suite.[46] IboView,[47] Molden,[48] and GaussView[49] software packages were used to produce molecular orbitals.

## III. Results and discussion

### IIIA. M(Tren)$_{1,2}$, M(Azacrypt), and M(Tripip) (M = Na, K) complexes

The polarization of valence $ns^1$ electron of M (M = Na, K) facilitates an efficient dative attack from the Tren ligand by minimizing the metal-ligand electron-electron repulsion (see the molecular orbital of Na(Tren) in Figure 1). The chelation of M(Tren) with another Tren ligand forces the valence electron to the periphery of the complex (compare the orbitals of Na(Tren) and Na(Tren)$_2$ in Figure 1). Na(Tren) can be recognized as "quasi-valence-type" based on the high electron localization on the metal center, whereas Na(Tren)$_2$ is a "surface-type" complex as a result of its distributed peripheral electron cloud (see Refs. 50 and 51 for more information on this terminology). In ground states, all studied M(Tren)$_2$, M(Azacrypt), and M(Tripip) (M = Na, K) systems carry an expanded electron cloud and hence they belong to the surface-type category. An accurate representation of this diffuse electron cloud is vital for analyzing their geometries.[23,29,30] Specifically, the implemented CAM-B3LYP method with a diffuse basis set in this work has been proven to produce accurate geometries on them.[52] Surface-type structures are known to have longer N−H and shorter M−N bonds due to the N$^{\delta-}$−H$^{\delta+}$ and M$^{\delta+}$−N$^{\delta-}$ charge localization. Specifically, the exterior electron cloud attracts the $\delta+$ charged H−atoms which stretches N−H bonds whereas its repulsion with the $\delta-$ charged N−atoms compresses the M−N bonds (see Refs. 28 and 53). These patterns are preserved by the systems described in this work (compare the slightly longer N−H and shorter M−N$_e$ (N$_e$ = equatorial N-atoms) bonds of neutral systems compared to their cations listed in Table 1). Note that the interaction of N-atoms in the equatorial positions with the doughnut-shaped diffuse electron cloud is greater compared to the axial N-atoms (Figure 1) and hence the pattern is not maintained in M−N$_a$ (N$_a$ = axial N−atoms) bonds (Table 1).

Generally, as we move down of the periodic table the D$_e$s of alkali metal-complexes decrease.[23,54,55] This is in line with all the studied systems in this work except for Na(Azacrypt) which has a 4.47 kcal/mol lower D$_e$ compared to K(Azacrypt) (Table 1). The total number of coordination sites of each M(Tren)$_2$, M(Azacrypt), and M(Tripip) complex is eight and their D$_e$s are comparable. Among the neutral complexes, the largest D$_e$s were observed for M(Tren)$_2$ whereas the comparatively lesser D$_e$s of M(Azacrypt) and M(Tripip) may be a result of the combination of the higher stress of bonds in their cage-type structures and the diffuse electron cloud. Upon comparing the cage-type structures, the Na(Tripip)$^{0,+}$ show higher $D_e$s compared to Na(Azacrypt)$^{0,+}$ whereas the opposite is true for K(Tripip)$^{0,+}$ vs. K(Azacrypt)$^{0,+}$. In all cases the cationic complexes carry larger $D_e$s compared to the neutral complexes owing to the stronger electrostatic attraction between M$^+$ and electron rich ligands.

The IEs of surface-type complexes are lower compared to the quasi-valence-type structures.[30,56,57] This is further corroborated by this work; i.e., the AIEs and VIEs (vertical IEs) of M(Tren)$_2$ are 0.5–0.6 eV smaller compared to the M(Tren) at all levels of theory (Table 1). Note that the D2, P3, and P3+ VIEs were obtained by calculating −VEAEs of the cationic systems using the geometries of their neutral structures. VIEs obtained at D2, P3, and P3+ levels agree within 0–0.007 eV. The IEs decrease in the order of M(Tren)$_2$ > M(Azacrypt) > M(Tripip) and the pattern strongly correlates to the number of N−H terminals of the complexes rather than the number of C−H terminals. Specifically, going from M(Tren)$_2$ to M(Azacrypt) the drop of IE per N−H vs. C−H is 0.054 vs. 0.027 eV and that for the M(Azacrypt) → M(Tripip) is 0.013 vs. 0.007 eV. This highlights the importance of N−H terminals on the stabilization of the diffuse electron cloud against the ionization. All studied Na-complexes carry larger IEs compared to their corresponding K-complexes. Notice that the IEs of Na (5.139 eV) and K (4.341 eV) atoms are more than twice as high compared to the IEs of the corresponding complexes.[7] Indeed, the IEs of all studied complexes are smaller than that of any atom in the periodic table and hence can be recognized as "superalkalis".[31] The chelating ligand of the previously studied Na[2.2.2]cryptand has a denticity of eight similar to Azacrypt and Tripip ligands, and the structure of Na[2.2.2]cryptand closely resembles the Na(Azacrypt) where the O−atoms of the former is replaced by N−H in the latter.[23,36] The reported D2 VIE of Na[2.2.2]cryptand is 1.669 eV which is lower by 0.099 eV from value of the Na(Azacrypt) and interestingly closer to that of K(Tripip); i.e., 1.694 eV.

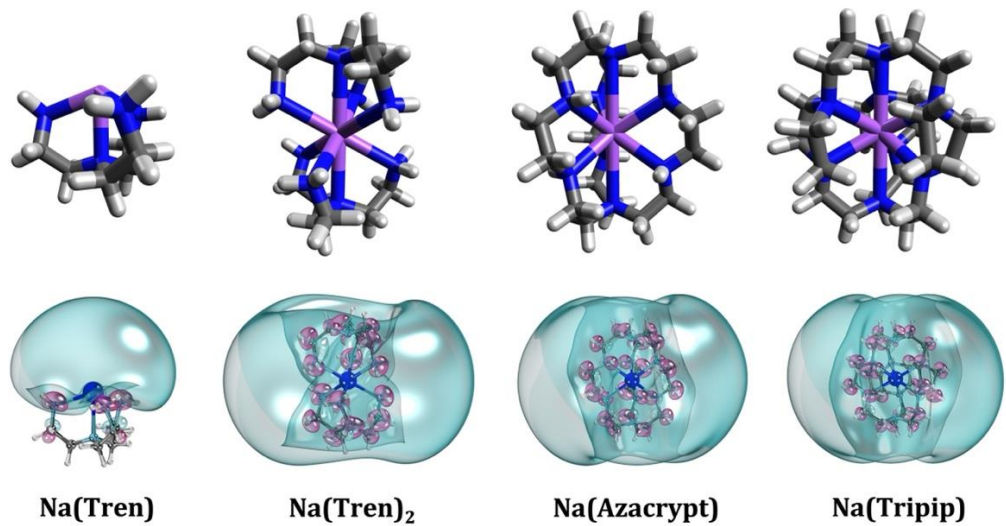

Figure 1. Geometries (top-row) and the contours of singly occupied molecular orbitals (bottom-row) of Na(Tren)$_{n=1,2}$, Na(Azacrypt), and Na(Tripip) complexes. A threshold of 70% was used. Structures and the orbital contours of corresponding potassium complexes have similar shapes.

Table 1. Point group symmetries, dissociation energies ($D_e$, kcal/mol), adiabatic ionization energies (AIE, eV), vertical ionization energies (VIE, eV), and equilibrium M−N and N−H bond distances ($r_e$, Å) of M(L)$_n^{0,+}$ systems.

| Species | Point group | $D_e$ | AIE | VIE | | | $r_e$ [a] | | |
|---|---|---|---|---|---|---|---|---|---|
| | | | | D2 | P3 | P3+ | M−N$_e$ | M−N$_a$ | N−H |
| Na(Tren) | C$_3$ | 21.33 | 2.689 | 2.613 | 2.618 | 2.618 | 2.460 | 2.640 | 1.019 |
| Na(Tren)$^+$ | C$_3$ | 82.98 | | | | | 2.432 | 2.421 | 1.018 |
| K(Tren) | C$_3$ | 17.66 | 2.655 | 2.512 | 2.519 | 2.518 | 2.871 | 3.099 | 1.018 |
| K(Tren)$^+$ | C$_3$ | 58.19 | | | | | 2.831 | 2.849 | 1.018 |
| Na(Tren)$_2$ | D$_3$ | 33.24 | 2.184 | 2.008 | 2.012 | 2.011 | 2.639 | 3.148 | 1.019 |
| Na(Tren)$_2^+$ | D$_3$ | 106.53 | | | | | 2.663 | 3.068 | 1.017 |
| K(Tren)$_2$ | D$_3$ | 33.12 | 2.122 | 1.953 | 1.959 | 1.959 | 2.950 | 3.205 | 1.019 |
| K(Tren)$_2^+$ | D$_3$ | 85.93 | | | | | 2.975 | 2.160 | 1.017 |
| Na(Azacrypt) | D$_3$ | 22.74 | 1.863 | 1.768 | 1.767 | 1.767 | 2.769 | 3.343 | 1.019 |
| Na(Azacrypt)$^+$ | D$_3$ | 103.43 | | | | | 2.771 | 3.362 | 1.019 |
| K(Azacrypt) | D$_3$ | 27.21 | 1.841 | 1.751 | 1.751 | 1.751 | 2.946 | 3.180 | 1.019 |
| K(Azacrypt)$^+$ | D$_3$ | 86.51 | | | | | 2.965 | 3.191 | 1.018 |
| Na(Tripip) | D$_3$ | 25.89 | 1.783 | 1.694 | | | 2.678 | 3.099 | |
| Na(Tripip)$^+$ | D$_3$ | 108.45 | | | | | 2.681 | 3.095 | |
| K(Tripip) | D$_3$ | 23.81 | 1.733 | 1.680 | | | 2.846 | 2.984 | |
| K(Tripip)$^+$ | D$_3$ | 83.29 | | | | | 2.884 | 2.988 | |

[a]N$_e$ and N$_a$ represent equatorial and axial N−atoms, respectively.

Excited state analysis of highly diffuse complexes is known to be associated with great challenges.[57–59] Especially, the predictions are sensitive to the quality of the utilizing level of theory and the basis sets. The EPT methods are proven to provide rather precise excitation energies for such systems.[30,52,58] The accuracy and computational cost of EPT methods implemented in this work increases in the order of the KT < D2 < P3 < P3+. Indeed, the predictions of D2, P3, and P3+ methods on the excitation energies of highly diffuse systems are in great harmony (agree within 0.1 eV) with the state-of-art EOM-EA-CCSD (equation of motion coupled cluster singles and doubles method based on electron attachment) and CASPT2 (complete active space second-order perturbation theory).[30,58,60] Note that in this work only EPT methods were implemented for excited state analysis since EOM-EA-CCSD and CASPT2 calculations with larger basis sets for such systems are rather challenging due to extreme computational demand. Our past benchmark studies support the hypothesis that the application of d-aug-cc-pVXZ (X = D, T) on the terminal H–atoms that directly interact with the diffuse electron cloud is crucial (see Refs. 52 and 53). On the other hand, application of cc-pVXZ (D, T) on the inner atoms is sufficient as an accuracy-efficiency compromise.[52,53] In this work cc-pVDZ (M, C, N) d-aug-cc-pVDZ (H) basis set was selected to analyze excited states of all surface-type complexes.

Figure 2 provides the P3+ vertical excitation energies and the occupying Dyson orbitals of Na(Tren)$_2$. The singly occupied doughnut-shaped molecular orbital in the ground state is an approximate to a SA s-orbital (1S). In excited states this electron populates p-, d-, f-, and g-type orbitals (or SA P-, D-, F-, G-orbitals). Under D$_3$ symmetry the SA P-, D-, F-, and G-configurations split into {E and A$_2$}, {2E, A$_1$}, {2E, A$_1$, 2A$_2$}, and {3E, A$_1$, 2A$_2$} components, respectively. The first two electron transitions, 0.162 eV and 0.549 eV, correspond to SA 1P$^1$ configurations. The proceeding three excited states carry SA 1D$^1$ configurations (at 0.583, 0.744, and 0.760 eV). The SA 2S$^1$ (at 1.131 eV) lies among the splits of 1F$^1$ (1.078–1.167 eV). The second set of SA P-level populates at 1.274 eV and is followed by 1G$^1$ set. In overall, the observed SA shell-model is 1S, 1P, 1D, 1F, 2S, 2P, and 1G. The calculated KT, D2, P3, and P3+ excitation energies of all M(Tren)$_2$, M(Azacrypt), and M(Tripip) (M = Na, K) are listed in Table 2. See SI for more information on their excitation energies. The excited states of K(Tren)$_2$ were also analyzed up to the 1G level and the shell model remains the same. All excitation energies of K(Tren)$_2$ are slightly lower compared to Na(Tren)$_2$ (by 0–0.070 eV), except for the 1P$^1$ (1$^2$E) state which is higher by only 0.002 eV (Tables S15 and S17). In this system only one component of 2P$^1$ ($^2$E) (at 1.224 eV) was identified as similar to Na(Tren)$_2$. Notice that the excited states of M(Azacrypt) and M(Tripip) (M = Na, K) were studied only up to 1F level due to the high computational cost. Comparatively, the computational cost of M(Tripip) is greater and hence only D2 level of theory was adopted. As a trend, excitation energies decrease moving from M(Tren)$_2$ to M(Azacrypt) and to M(Tripip) (compare values of Tables S15, S17, S19, S21, S23, and S25). Slightly lower excitation energies were observed for K(Azacrypt)/K(Tripip) (by 0–0.03 eV) compared to their Na-complexes.

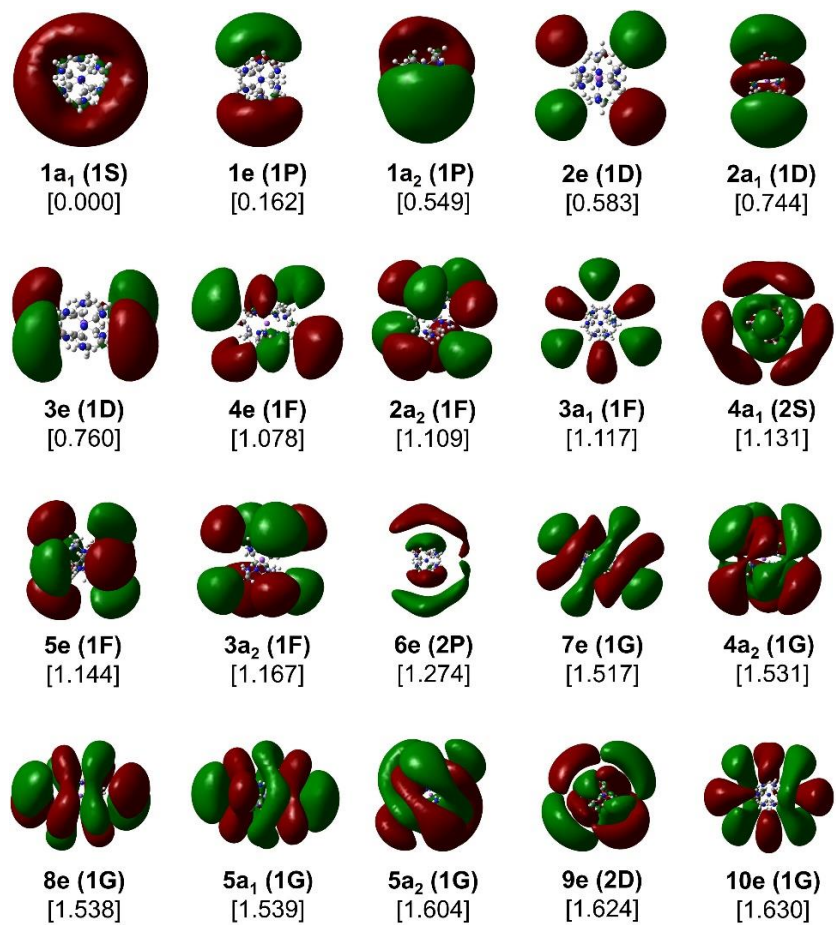

Figure 2. Representative Dyson orbitals for vertical electron attachment of Na(Tren)$_2$$^+$ (D$_3$ point group). In each case the approximate SA shell is given in parenthesis. The P3+ vertical excitation energies (eV) are listed in square brackets.

Table 2. Vertical excitation energies (eV) of M(Tren)$_2$, M(Azacrypt), and M(Tripip) (M = Na, K) at EPT.

| Approximate SA shell | KT | D2 | P3 | P3+ |
|---|---|---|---|---|
| | | Na(Tren)$_2$ | | |
| 1S | 0.000 | 0.000 | 0.000 | 0.000 |
| 1P | 0.122, 0.393 | 0.163, 0.553 | 0.162, 0.549 | 0.162, 0.549 |
| 1D | 0.458–0.586 | 0.586–0.764 | 0.584–0.760 | 0.583–0.760 |
| 1F | 0.883–0.930 | 1.080–1.172 | 1.078–1.168 | 1.078–1.167 |
| 2S | 1.020 | 1.126 | 1.133 | 1.131 |
| 2P[a] | 1.136 | 1.271 | 1.275 | 1.274 |
| 1G | 1.274–1.379 | 1.518–1.631 | 1.518–1.631 | 1.517–1.630 |
| | | K(Tren)$_2$ | | |
| 1S | 0.000 | 0.000 | 0.000 | 0.000 |
| 1P | 0.124, 0.379 | 0.164, 0.518 | 0.163, 0.516 | 0.164, 0.517 |
| 1D | 0.445–0.568 | 0.570–0.722 | 0.568–0.720 | 0.569–0.721 |
| 1F | 0.861–0.906 | 1.043–1.124 | 1.042–1.122 | 1.042–1.122 |
| 2S | 0.975 | 1.057 | 1.064 | 1.064 |
| 2P[a] | 1.111 | 1.219 | 1.224 | 1.224 |
| 1G | 1.242–1.396 | 1.468–1.577 | 1.470–1.579 | 1.470–1.579 |
| | | Na(Azacrypt) | | |
| 1S | 0.000 | 0.000 | 0.000 | 0.000 |
| 1P | 0.055, 0.238 | 0.014, 0.312 | 0.020, 0.303 | 0.019, 0.304 |
| 1D | 0.381–0.450 | 0.464–0.550 | 0.458–0.543 | 0.459–0.543 |
| 1F | 0.709–0.772 | 0.811–0.933 | 0.811–0.924 | 0.811–0.925 |
| | | K(Azacrypt) | | |
| 1S | 0.000 | 0.000 | 0.000 | 0.000 |
| 1P | 0.045, 0.227 | 0.000, 0.298 | 0.006, 0.290 | 0.005, 0.291 |
| 1D | 0.380–0.437 | 0.460–0.533 | 0.455–0.527 | 0.455–0.527 |
| 1F | 0.695–0.759 | 0.803–0.914 | 0.802–0.907 | 0.802–0.907 |
| | | Na(Tripip) | | |
| 1S | | 0.000 | | |
| 1P | | 0.130, 0.223 | | |
| 1D | | 0.432–0.476 | | |
| 1F[b] | | 0.761–0.824 | | |
| | | K(Tripip) | | |
| 1S | | 0.000 | | |
| 1P | | 0.128, 0.212 | | |
| 1D | | 0.425–0.461 | | |
| 1F[b] | | 0.734–0.815 | | |

[a] Within the focused energy range only one component of 2P$^1$ (6$^2$E) was seen for Na(Tren)$_2$ and K(Tren)$_2$ (Tables S15 and S17). [b] Similarly, in the selected range only three components of 1F$^1$ (3$^2$A$_1$, 4$^2$E, and 2$^2$A$_2$) were observed for Na(Tripip) and K(Tripip) (Tables S23 and S25).

**IIIB. M(Tren)$_2$@M′, M(Azacrypt)@M′, and M(Tripip)@M′ (M/M′ = Na, K) units**

The diffuse electron of M(L)$_n$ transfers to an incoming metal atom to produce a charge separated [M(L)$_n$]$^+$@M′$^-$. The D$_e^*$s with respect to [M(L)$_n$@M′]$^{0,+}$ → [M(L)$_n$]$^{0,+}$ + M′ fragmentation of studied species are listed in Table 3. The D$_e^*$s of neutral complexes decrease in the order of M(Azacrypt)@M′ > M(Tripip)@M′ > M(Tren)$_2$@M′ for every studied metal atom combination. Similarly, the D$_e^*$s varied in the series of K(L)$_n$@Na > Na(L)$_n$@Na > K(L)$_n$@K > Na(L)$_n$@K for all utilized ligand (L) types. Comparatively, the D$_e^*$s of cationic complexes are significantly lower. The MP2 D$_e^*$s were calculated only for neutral and cationic Na(L)$_2$@Na, Na(L)$_2$@K, K(L)$_2$@Na for L = Tren and their discrepancies with corresponding CAM-B3LYP values are less than 1.2 kcal/mol. As expected, the lower binding energy between [M(L)$_n$]$^+$ + M′ correlates to longer M···M distances of [M(L)$_n$@M′]$^+$; i.e., longer by 0.8–1.8 Å compared to their neutral units (Table 3). The largest ΔE$_{S-T}$ were observed for the M(Azacrypt)@M′ units, and they are higher for M′ = Na compared to the M′ = K cases (Table S29).

Going one step further the anions of Na(L)$_n$@Na were also studied at the D2 level of theory. The −VEAEs are reported in the Table 3. Note that there are two −VEAEs per each focused system which correspond to the ground and the first excited state of the anions. The computed electron affinities varied in the series of Na(Azacrypt)@Na < Na(Tripip)@Na < Na(Tren)$_2$@Na. In the ground state the radical electron of [Na(L)$_n$@Na]$^-$ (1$^2$A) occupies a polarized Rydberg-like orbital (~SA 1S) and its promotion to a SA 1P-orbital produces the first excited state (2$^2$A) (Figure 3).

The lowest AIEs were observed for M(Tripip)@M′ and the values are 0.6–1.0 eV higher compared to the respective M(Tripip) complexes (compare corresponding AIEs listed in Tables 1 and 3). The greater AIEs of M(L)$_n$@M′ for M′ = Na compared to M′ = K correlates to the higher atomic electron affinity of the Na (0.548 eV) compared to K (0.501 eV).[7] The MP2 AIEs are 0.1–0.3 eV lower compared to the CAM-B3LYP values. Indeed, according to Boldyrev's definition these alkalide units too are superalkalis owing to their smaller IEs.[31,32] The structures of cations bear [M(L)$_n$]$^+$@M′ form, where one radical electron is present at the exterior metal atom (M′). In the first three excited states this electron populates the atomic p-orbitals of M′. The energies related to these transitions and occupying orbitals are given in Table S30 and Figure S1, respectively. Since the electron transition is almost perfect s → p of M′, the excitation energies of cationic units are expected to be comparable to the values of the bare M′–atoms. Indeed, the first three excitation energies of all M′ = Na and M′ = K cationic systems span within 1.992–2.160 and 1.425–1.591 eV, respectively, which are closer to the first excitation energies Na (2.10 eV) and K (1.61 eV).[61]

Table 3. Dissociation energies ($D_e^*$, kcal/mol), equilibrium M···M distances ($r_e$, Å), negative of vertical electron attachment energies (–VEAE, eV), and adiabatic ionization energies (AIE, eV) of $[M(L)_n@M']^{0,+}$ systems.

| Species | $D_e^{*a}$ | $r_e$ | –VEAE | AIE[a] |
|---|---|---|---|---|
| Na(Tren)$_2$@Na | 20.29 (19.11) | 5.500 | 0.411, 0.118 | 2.966 (2.691) |
| [Na(Tren)$_2$@Na]$^+$ | 2.28 (2.74) | 6.357 | | |
| Na(Tren)$_2$@K | 14.90 (14.28) | 6.073 | | 2.695 (2.431) |
| [Na(Tren)$_2$@K]$^+$ | 3.13 (3.91) | 6.881 | | |
| K(Tren)$_2$@Na | 21.35 (22.4) | 4.791 | | 2.955 (2.846) |
| [K(Tren)$_2$@Na]$^+$ | 2.13 (1.42) | 6.548 | | |
| K(Tren)$_2$@K | 15.40 | 5.806 | | 2.661 |
| [K(Tren)$_2$@K]$^+$ | 2.97 | 7.066 | | |
| Na(Azacrypt)@Na | 27.49 | 4.771 | 0.240, 0.101 | 2.973 |
| [Na(Azacrypt)@Na]$^+$ | 1.90 | 6.038 | | |
| Na(Azacrypt)@K | 21.34 | 5.501 | | 2.667 |
| [Na(Azacrypt)@K]$^+$ | 2.80 | 6.559 | | |
| K(Azacrypt)@Na | 27.62 | 4.968 | | 2.958 |
| [K(Azacrypt)@Na]$^+$ | 1.86 | 6.177 | | |
| K(Azacrypt)@K | 21.45 | 5.699 | | 2.653 |
| [K(Azacrypt)@K]$^+$ | 2.73 | 6.662 | | |
| Na(Tripip)@Na | 21.26 | 6.266 | 0.332, 0.135 | 2.643 |
| [Na(Tripip)@Na]$^+$ | 1.41 | 7.158 | | |
| Na(Tripip)@K | 16.39 | 6.848 | | 2.412 |
| [Na(Tripip)@K]$^+$ | 1.87 | 7.697 | | |
| K(Tripip)@Na | 21.81 | 6.107 | | 2.663 |
| [K(Tripip)@Na]$^+$ | 2.68 | 7.152 | | |
| K(Tripip)@K | 16.70 | 6.755 | | 2.421 |
| [K(Tripip)@K]$^+$ | 3.15 | 7.712 | | |

[a] MP2 $D_e^*$s and AIEs are given in parenthesis, whereas all other values obtained under CAM-B3LYP.

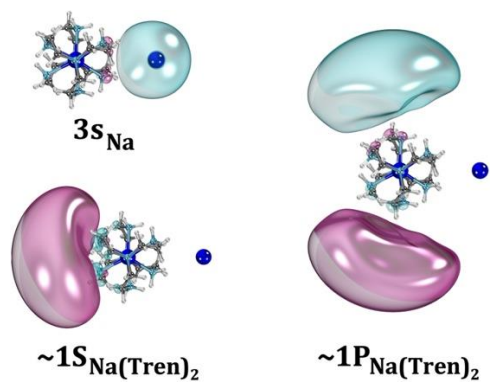

Figure 3. Selected orbitals of [Na(Tren)$_2$@Na]$^-$. The ~1S$_{Na(Tren)_2}$ and ~1P$_{Na(Tren)_2}$ orbitals are singly occupied in 1$^2$A and 2$^2$A states of the [Na(Tren)$_2$@Na]$^-$, respectively. The 3s$_{Na}$ is doubly occupied in both states. Orbitals of 1$^2$A and 2$^2$A of [Na(Azacrypt)@Na]$^-$ and [Na(Tripip)@Na]$^-$ have similar shapes.

## IV. Conclusions

The ground and electronically excited states of M(L)$_n$ (M = Na, K; L = Tren, Azacryptand, TriPip222; $n$ = 1, 2) molecular electrides and their analogous alkalide complexes, M(L)$_n$@M′, were analyzed using *ab initio* methods. The complexes are real minima on the potential energy surface, with positive D$_e$s with respect to their corresponding M(L)$_n$ → M + $n$L and M(L)$_n$@M′ → M(L)$_n$ + M′ dissociations. In the ground state M(L)$_n$ carries a diffuse electron in the periphery which occupies atomic-type orbitals in excited states. Consequently, they were identified as superatoms. The superatomic (SA) shell model discovered for M(Tren)$_2$ is 1S, 1P, 1D, 1F, 2S, 2P, and 1G. This orbital order is similar to the trends we have seen in the past for similar diffuse systems. Due to the high computational cost the excited states of M(Azacryptand) and M(TriPip222) were studied only up to the 1F level, and the order is consistent with that of M(Tren)$_2$. The excitation energies and ionization energies decreased moving from M(Tren)$_2$ to M(Azacryptand) and to M(TriPip222), and the trend correlates to the number of N−H terminals in them. The reaction between M(L)$_n$ and M′ causes an electron transfer from the former to the latter producing a charge separated structure for alkalide units; i.e., [M(L)$_n$]$^+$@M′$^-$ (M′ carries two electrons in the valence $n$s-orbital). The D$_e^*$s of M(L)$_n$@M′ decreased in the sequence of M(Azacryptand)@M′ > M(TriPip222)@M′ > M(Tren)$_2$@M′ (for all M and M′ combinations) and K(L)$_n$@Na > Na(L)$_n$@Na > K(L)$_n$@K > Na(L)$_n$@K (for all L). The radical electron occupies the valence $n$s-orbital of M′ of [M(L)$_n$@M′]$^+$ which gets promoted to its $n$p- and ($n$+1)s-orbitals in excited states. Similar to M(L)$_n$, in [Na(L)$_n$@Na′]$^-$ the radical electron occupies diffuse SA 1S- and 1P-orbitals in the ground and the first excited state.

**Conflicts of interest**

There are no conflicts to declare.

**Acknowledgements**

**TOC Graphics**

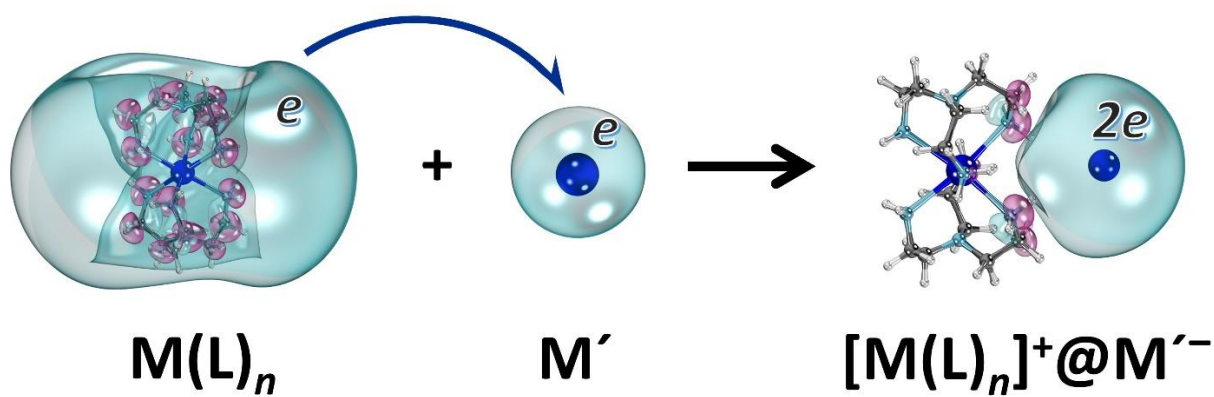